\documentclass[preprint,12pt]{elsarticle}

\usepackage{graphicx}
\usepackage{subfigure}
\usepackage{dcolumn}
\usepackage{bm}
\usepackage{multirow}
\usepackage{mathrsfs}
\usepackage{xcolor}

\usepackage{booktabs,array,makecell}
\newcolumntype{N}{@{}m{0pt}@{}}
\newcommand\ExtraSep
{\dimexpr\cmidrulewidth+\aboverulesep+\belowrulesep\relax}

\usepackage{amssymb}

\usepackage{braket}
\usepackage{amsmath}
\usepackage{amsfonts}

\usepackage{csquotes}

\usepackage[nolist]{acronym}
\usepackage{xcolor}

\usepackage{lscape}

\newcommand{\bea}{\begin{eqnarray}}
\newcommand{\eea}{\end{eqnarray}}
\newcommand{\be}{\begin{equation}}
\newcommand{\ee}{\end{equation}}

\newcommand{\Parab}{\text{D}}
\newcommand{\Tricomi}{\text{U}}
\newcommand{\Kummer}{\text{M}}

\newcommand{\Rmax}{\text{R}_{\max}}
\newcommand{\Iquad}{\mathcal{I}^\text{quad}}
\newcommand{\IcGTOs}{\mathcal{I}^\text{cGTOs}}
\newcommand{\Err}{\mathcal{E}_{\text{cGTOs}}}
\newcommand{\Jquad}{\mathcal{J}^\text{quad}}
\newcommand{\JcGTOs}{\mathcal{J}^\text{cGTOs}}
\newcommand{\Frr}{\mathcal{F}_{\text{cGTOs}}}

\journal{Advances in Quantum Chemistry}

\begin{document}

\begin{frontmatter}



\title{On the use of complex GTOs for the evaluation of radial integrals involving oscillating functions
}


\author{Abdallah Ammar$^{1,*}$, Arnaud Leclerc$^1$ and Lorenzo Ugo Ancarani$^{1,+}$}

\address{$^1$Universit\'e de Lorraine, CNRS, Laboratoire de Physique et Chimie Th\'eoriques,
	57000 Metz, France}

\address{$^*$Present postal address:  Université de Toulouse, CNRS,
Laboratoire de Chimie et Physique Quantiques, 31062 Toulouse, France}

\address{$^{+}$Corresponding author: email address: ugo.ancarani@univ-lorraine.fr}

\begin{abstract}
We study two classes of radial integrals involving a product of bound and continuum one-electron states.
Using a representation of the continuum part with an expansion on complex Gaussian Type Orbitals, such integrals can be performed analytically.
We investigate the reliability of this scheme for low-energy physical parameters. This study serves as a premise in view of potential applications in molecular scattering processes.
\end{abstract}


\begin{acronym}
        \acro{TDCS}{Triple Differential Cross Section}
        \acro{STOs}{Slater Type Orbitals}
        \acro{GTOs}{Gaussian Type Orbitals}
        \acro{cGTOs}{complex Gaussian Type Orbitals}
        \acro{rGTOs}{real Gaussian Type Orbitals}
	\acro{FBA}{First Born Approximation}
\end{acronym}

\end{frontmatter}


\section{Introduction}

The evaluation of integrals is a key issue in many atomic and molecular physics applications.
Multicenter molecular integrals involving one or more electrons bound states are ubiquitous and essential in quantum chemistry calculations.
The literature on the subject is abundant whether the bound functions involved are represented in terms of \ac{STOs} or \ac{GTOs}. The latter have
become particularly popular because a number of mathematical properties allows an efficient evaluation of multielectron bound states integrals \cite{boys_1950,hill_2013}.

When one or several electrons are in a continuum state, the integrals are more difficult and the integration tools developed for bound states
are not necessarily adapted, especially in the molecular case. The present manuscript is dedicated to a class of integrals involving products of one-electron
bound and continuum states. An analytical approach based on a \ac{cGTOs} representation of the continuum is proposed and its reliability is numerically investigated. While the
ultimate goal would be to reach an all-Gaussian approach to evaluate the required multicentric and multielectrons integrals involved in scattering processes, we
start here with a monocentric study and one-electron functions.
The present work aims to better grasp the potential, the numerical efficiency and the range of applicability of the proposed strategy.

In the study of the ionization of atoms or molecules by a projectile, an electron ($\mathbf{r}$) initially in a bound state $\phi_I$ is ejected into a continuum state
$\psi^-_{\mathbf{k_e}}$ with a given momentum $\mathbf{k_e}$. To calculate measurable quantities such as cross sections,
within the first Born approximation one encounters the following matrix element (see, \textit{e.g.}, \cite{Bransden_2003}): 
\begin{equation}
	\Braket{\psi^-_{\mathbf{k_e}} \left( \mathbf{r} \right)
			| e^{\imath \mathbf{q} \cdot \mathbf{r}}  |
			\phi_I\left( \mathbf{r} \right)}
	\text{,}
	\label{eq:T0}
\end{equation}
where $\mathbf{q}$ is the momentum transferred to the target by the projectile. For simplicity, here both electronic wave functions are written in a one-center description.
One also encounters the case $\mathbf{q}=\mathbf{0}$, that is to say the overlap $	\Braket{\psi^-_{\mathbf{k_e}} \left( \mathbf{r} \right) |
			\phi_I\left( \mathbf{r} \right)}$
which should vanish if the initial and final states are exact solutions of the same Hamiltonian.
However, this is rarely the case since the two wave functions are usually calculated with different methods and making different approximations.


The factor $e^{\imath \mathbf{q} \cdot \mathbf{r}}$ recalls obviously also the Fourier transform.
Since some seminal papers such as   \cite{franklin_1962, Bonham_1964},
the Fourier transform method has been widely used in quantum chemistry. The integrals that appear, though, involve only bound states.
Should Fourier transform techniques be envisaged for applications in which a combination of bound and continuum states is involved, matrix elements such as  (\ref{eq:T0}) would appear.

By using the Rayleigh expansion for $e^{\imath \mathbf{q} \cdot \mathbf{r}}$ and the standard expansions in spherical coordinates
for the initial and final states, the angular integrations can be treated separately and performed analytically. One is thus left with the evaluation of radial
integrals which is a challenge because the integrand oscillates up to large distances.

The purpose of this manuscript is twofold. First, we wish to explore numerical issues related to the different parameters, in particular
the momenta $k_e$ and $q$ that dictate the oscillations of the integrand, as well as the effective infinite radial distance to
be considered (it depends essentially on the extension of the initial state).
The second purpose is to test, through the evaluation of radial integrals, the efficiency and reliability of representing radial continuum functions by a
finite number $N$ of \ac{cGTOs}. The idea behind this approach is to reach a way to evaluate matrix elements exploiting the mathematical
properties of GTOs~\cite{ammar_2020,ammar_2021_I,ammar_2021_II}.
This will be particularly useful in molecular applications for which computationally expensive multicentric integrals have to be calculated.
In the present numerical investigation, we shall limit ourselves to one$-$center problems with the intention of putting our proposal on solid grounds.

Section 2 introduces the radial integrals investigated here and provides their analytical evaluation through the \ac{cGTOs} representation of the continuum states.
The numerical investigation is presented in Section 3. First the integrand and the range of integration is examined for different sets of parameters.
Then, the integrals evaluated analytically and numerically are compared.
A summary is given in Section 4.

\section{Theoretical formulation}
\label{sec:Theory}

\subsection{Radial integrals}

Whether the electron in the continuum is described by a plane wave, a Coulomb or a distorted wave, we use the standard partial wave expansion
\begin{equation}
	\psi^-_{\mathbf{k_e}}(\mathbf{r}) = \sqrt{\frac{2}{\pi}}
	\sum_{l=0}^{\infty} \sum_{m=-l}^l (\imath)^l
	e^{-\imath \delta_l }
	\frac{u_{l,k_e}(r)}{k_e r} Y_l^m(\hat{r}) Y_l^{m*}(\widehat{k_e})
	\text{,}
	\label{eq:psi_f}
\end{equation}
where $\delta_l$ denotes the phase shift for a given angular momentum $l$ and $Y_{l}^{m}$ the complex spherical harmonics. The radial functions $u_{l,k_e}(r)$  are the solutions of the ordinary differential equation
\begin{equation}
        \left[ -\frac{1}{2} \frac{d^2}{dr^2} + \frac{l(l+1)}{2r^2}
        + U(r)  \right] u_{l,k_e}(r)
        = \frac{k_e^2}{2} \, u_{l,k_e}(r)
        \text{,}
        \label{eq:fct_disto_DE}
\end{equation}
where $U(r)$ is the potential felt by the ejected electron.
We also make use of the Rayleigh expansion~\cite{magnus_2013}
\begin{equation}
	e^{\imath \mathbf{q} \cdot \mathbf{r}}  =
	4 \pi \sum_{\lambda=0}^{\infty}\sum_{\mu=-\lambda}^{\lambda}  \,
	\imath^{\lambda} \,  j_{\lambda}\left( qr \right) \,
	Y_{\lambda}^{\mu*}\left( \hat{q} \right)
	Y_{\lambda}^{\mu}\left( \hat{r} \right)
	\text{,}
\end{equation}
which involves the spherical Bessel function $j_{\lambda}$.

We consider here an initial state $\phi_I \left( \mathbf{r} \right)$ centred on an atomic nucleus or on the heaviest
nucleus ($\mathbf{R}=\mathbf{0}$) in the case of a polyatomic molecule with a heavy center.
In a standard partial wave expansion, we consider a linear combination of $N_I$ terms
\begin{equation}
	\phi_I \left( \mathbf{r} \right)  = \sum_{i=1}^{N_I} C_{i} \, R_i(r) \, Y_{l_i}^{m_i} \left( \hat{r} \right)
	\text{.}
	\label{eq:MO_def}
\end{equation}
The radial part $R_i(r)$ considered here are either \ac{STOs} $R_i(r) = r^{n_i-1} \, e^{-\zeta_i r}$
or \ac{GTOs} $R_i(r)=r^{n_i-1} \, e^{-\gamma_i r^2}$ where $n_i$ are strictly positive integers, $\zeta_i$ and $\gamma_i$ are real parameters.

To evaluate matrix elements such as (\ref{eq:T0}), all angular parts are treated separately with standard techniques \cite{Edmonds_1996}, and one is left with
radial integrals.  Two families appear, depending on whether one considers \ac{STOs} or \ac{GTOs}:
\begin{equation}
	\mathcal{I}_{l,k_e,\lambda}\left( \zeta, n, q \right) =
	\int_0^{\infty} \left(u_{l,k_e}(r)\right)^*  \, r^{n} e^{-\zeta r} \, j_{\lambda}\left( qr \right)  d {r}
	\text{,}
	\label{eq:RadInteg_Idef}
\end{equation}
\begin{equation}
	\mathcal{J}_{l,k_e,\lambda}\left( \gamma, n, q \right) =
	\int_0^{\infty} \left(u_{l,k_e}(r)\right)^*  \, r^{n} e^{-\gamma r^2} \, j_{\lambda}\left( qr \right)  d {r}
	\text{.}
	\label{eq:RadInteg_Jdef}
\end{equation}
The particularity, here, is that the radial functions $u_{l,k_e}$  are associated to continuum states, and thus they
oscillate up to infinity, and more so as $k_e$ increases.
The integrand of either $\mathcal{I}_{l,k_e,\lambda}$ or $\mathcal{J}_{l,k_e,\lambda}$ involves also another oscillating
function, the Bessel function $j_{\lambda}\left( qr \right)$. As a result, depending on the values of $k_e$, $q$, $\lambda$
and $l$, the evaluation of the integrals may not be easy.
In spite of the existence of highly accurate one$-$dimensional integration libraries to evaluate
this kind of integrals, an analytical scheme based on the \ac{GTOs} representation of the radial functions $u_{l,k_e}$ could be more suitable.
In fact, such a strategy is put forward since it should be particularly valuable when evaluating two$-$electron integrals
that appear in molecular calculation where direct numerical integration will be computationally very expensive. We mention that a first study of one-electron integrals but with multicentric  bound states $\phi_I \left( \mathbf{r} - \mathbf{R}\right)$ with $\mathbf{R}\ne \mathbf{0}$  has been presented in \cite{ammar_2021_II}.

Special subcases of such integrals are obtained when the Bessel function is absent, that is to say when $q=0$.
All integrals vanish except when $\lambda=0$, and we denote the radial integrals as
\begin{equation}
	{\mathcal{\tilde I}}_{l,k_e}\left( \zeta, n \right) =
	\int_0^{\infty} \left(u_{l,k_e}(r)\right)^*  \, r^{n} e^{-\zeta r} \,  d {r}
	\text{,}
	\label{eq:RadInteg_Idef_q0}
\end{equation}
\begin{equation}
	{\mathcal{\tilde J}}_{l,k_e}\left( \gamma, n \right) =
	\int_0^{\infty} \left(u_{l,k_e}(r)\right)^*  \, r^{n} e^{-\gamma r^2}  d {r}
	\text{.}
	\label{eq:RadInteg_Jdef_q0}
\end{equation}
Note that if the radial functions $u_{l,k_e}(r)$ were those corresponding to a bound state, such integrals are the
standard one$-$electron integrals well documented in the atomic physics or quantum chemistry literature.

\subsection{\ac{cGTOs} representation of radial continuum functions}

In ref.~\cite{ammar_2020,ammar_2021_I,ammar_2021_II} we have proposed to employ \ac{cGTOs} to represent radial
continuum functions defined by the expansion
\begin{equation}
	u_{l,k_e}(r) \approx r^{l+1} \, \sum_{s=1}^{N} \,
		\left[ c_s \right]_{l,k_e}  \, e^{-\left[ \alpha_s \right]_{l} r^2}.
	\label{eq:cG_def}
\end{equation}
The exponents and the coefficients of \ac{cGTOs} are defined in the complex plane $c_s, \alpha_s \in \mathbb{C}$,
 the real part of exponents $\Re(\alpha_s)$ being constrained to be positive as to ensure square integrability.
The \ac{cGTOs} combination is multiplied by $r^{l+1}$
in order to reproduce the expected behaviour of $u_{l,k_e}(r)$ at small radial distance $r$.
For a given $l$, we employ a fixed set of $N$ exponents to represent a set of radial functions $\{u_{l,k_e}\}$;
the $N$ linear coefficients $\left[ c_s \right]_{l,k_e}$ are optimized
for each wave number $k_e$. The details of our numerical scheme, based on a non-linear optimization of the exponents alternating with a least-square optimization of the coefficients, are described in~\cite{ammar_2020}.

\subsection{Analytical form for the radial integrals}
\label{AnalyticalScheme}

Making use of the \ac{cGTOs} representation (\ref{eq:cG_def}), the integrals (\ref{eq:RadInteg_Idef}) and (\ref{eq:RadInteg_Jdef}) become
\begin{equation}
	\mathcal{I}_{l,k_e,\lambda}\left( \zeta, n, q \right) = \sum_{s=1}^{N}
		\left[ c_s \right]^*_{l,k_e}
	\int_0^{\infty} e^{-\left[ \alpha_s \right]^*_{l} r^2}  \, r^{n+l+1} e^{-\zeta r} \, j_{\lambda}\left( qr \right)  d {r}
	\text{,}
	\label{eq:RadInteg_IdefGTO}
\end{equation}
\begin{equation}
	\mathcal{J}_{l,k_e,\lambda}\left( \gamma, n, q \right) = \sum_{s=1}^{N}
		\left[ c_s \right]^*_{l,k_e}
	\int_0^{\infty} e^{-\left[ \alpha_s \right]^*_{l} r^2}  \, r^{n+l+1} e^{-\gamma r^2} \, j_{\lambda}\left( qr \right)  d {r}
	\text{.}
	\label{eq:RadInteg_JdefGTO}
\end{equation}
Each integrals in (\ref{eq:RadInteg_IdefGTO}) can be expressed analytically as a finite Hankel sum of
special functions. Details of the derivation are provided in \cite{ammar_2023},
and only the final results are given here:
\begin{equation}
	\mathcal{I}_{l,k_e,\lambda}\left( \zeta, n, q \right) =
		\sum_{s=1}^{N} \left[ c_s \right]_{l,k_e}^*
                \left[
                \mathcal{I}_{l,k_e,\lambda}^{(1)} \left( \zeta, n, q, \left[ \alpha_s \right]_l^*\right) +
                \mathcal{I}_{l,k_e,\lambda}^{(2)} \left( \zeta, n, q, \left[ \alpha_s \right]_l^*\right) \right]
        \text{,}
        \label{eq:I_analytical}
\end{equation}
with
\begin{equation}
\begin{aligned}
	\mathcal{I}_{l,k_e,\lambda}^{(1)} &\left( \zeta, n, q, \left[ \alpha_s \right]_l^*\right)
	= \frac{(-\imath)^{\lambda}}{2 \imath}
	\sum_{k=0}^{[\lambda/2]}
	\frac{(-1)^k a_{2k}\left( \lambda + \frac{1}{2} \right)}{q^{2k+1}} \\
	& \times \left[
	\mathcal{G} \left( \left[ \alpha_s \right]_l^*, \zeta - \imath q, n+l-2k \right)
	-(-1)^{\lambda}  \,
	\mathcal{G} \left( \left[ \alpha_s \right]_l^*, \zeta + \imath q, n+l-2k \right)
	\right]
        \text{,}
	\label{eq:I1_sum_def}
\end{aligned}
\end{equation}
\begin{equation}
\begin{aligned}
	\mathcal{I}_{l,k_e,\lambda}^{(2)} &\left( \zeta, n, q, \left[ \alpha_s \right]_l^*\right)
        = \frac{(-\imath)^{\lambda}}{2}
	\sum_{k=0}^{[(\lambda-1)/2]} \frac{(-1)^k a_{2k+1}\left( \lambda + \frac{1}{2} \right)}{q^{2k+2}} \\
	& \times \left[
	\mathcal{G} \left( \left[ \alpha_s \right]_l^*, \zeta - \imath q, n+l-2k-1 \right)
	+ (-1)^{\lambda}
	\mathcal{G} \left( \left[ \alpha_s \right]_l^*, \zeta + \imath q, n+l-2k-1 \right)
	\right]
        \text{,}
	\label{eq:I2_sum_def}
\end{aligned}
\end{equation}
where the square bracket $[x]$
in the upper bound of the summations denotes the integer part of $x$, and the sum is zero if the lower bound exceeds the upper bound.
Above, $\mathcal{G}$ stands for the integral
\begin{equation}
\begin{aligned}
	\mathcal{G} \left( a, b, \mu \right)
	&\equiv
	\int_0^{\infty} e^{-a r^2 - b r} \,  r^{\mu} \, dr \\
	&=
	\frac{ \Gamma\left(\mu+1\right) }{(2 a)^{\frac{\mu+1}{2}}} \,
	\exp \left( \frac{b^2}{8 a} \right) \,
	\Parab_{\mu+\frac{1}{2}}\left( \frac{b}{\sqrt{2 a}} \right) \\
	&=
	\frac{ \Gamma\left(\mu+1\right) }{(4 a)^{\frac{\mu+1}{2}}} \,
	\Tricomi \left(\frac{\mu+1}{2}, \frac{1}{2}, \frac{b^2}{4 a}\right)
        \text{,}
	\label{eq:G_def}
\end{aligned}
\end{equation}
expressed in terms of the special functions $\Parab_{\cdot}(\cdot)$ or $\Tricomi(\cdot,\cdot,\cdot)$
which are the parabolic cylinder function and the Tricomi confluent hypergeometric
function, respectively~\cite{magnus_2013,bateman_1953,gradshteyn_2007,olver_2010}.
While the result (\ref{eq:I_analytical}) is analytical, the evaluation of the sum of special functions is ultimately performed numerically.

The integrals~\eqref{eq:RadInteg_JdefGTO}  can be written in a
simpler closed form in terms of Kummer confluent hypergeometric function $\Kummer(\cdot,\cdot,\cdot)$~\cite{gradshteyn_2007}
\begin{equation}
\label{JKummer}
\begin{aligned}
	\mathcal{J}_{l,k_e,\lambda}&\left( \gamma, n, q \right)
	=
	\frac{\sqrt{\pi}}{4} \, \left(\frac{q}{2}\right)^{\lambda} \,
	\frac{\Gamma\left( \frac{l+\lambda+n+2}{2} \right) }
	     {\Gamma\left( \lambda + \frac{3}{2} \right)} \\
	& \quad \times \sum_{s=1}^{N}
	\frac{\left[ c_s \right]^*_{l,k_e} }
	     {\left(\left[ \alpha_s \right]^*_{l} + \gamma \right)^{\frac{l+\lambda+n+2}{2}}} \, \,
	\Kummer \left( \frac{l+\lambda+n+2}{2}, \lambda+\frac{3}{2}, \frac{-q^2}{4 \left( \left[ \alpha_s \right]^*_{l} + \gamma \right)} \right)
	\text{.}
\end{aligned}
\end{equation}

Both integrals~\eqref{eq:RadInteg_IdefGTO} and~\eqref{eq:RadInteg_JdefGTO} are always convergent. 
However, the finite Hankel series~\eqref{eq:I1_sum_def} and~\eqref{eq:I2_sum_def} may contain divergent terms
if $l+n-\lambda$ is a negative integer or equal to zero, due to the Gamma function.
This situation does not appear for physical parameters owing to the restriction imposed by the angular integrals.
If, for some reason, one should mathematically consider such situations, the singularities in the Hankel series can be removed by an $\epsilon-$expansion
technique~\cite{tomaschitz_2013}.

The present manuscript aims to investigate the applicatibility and reliability of the proposed analytical scheme for the evaluation of the two integrals $\mathcal{I}_{l,k_e,\lambda}\left( \zeta, n, q \right)$
and $\mathcal{J}_{l,k_e,\lambda}\left( \gamma, n, q \right)$.
The simpler cases of integrals ${\mathcal{\tilde I}}_{l,k_e}\left(\zeta, n \right)$ and ${\mathcal{\tilde I}}_{l,k_e}\left( \zeta, n \right)$ that correspond to $q=0$ (and thus $\lambda=0$) do not lead to extra valuable information and,
for the sake of space,
will not be scrutinized here.

\section{Numerical investigation}
\label{sec:Potentials}

In the investigation to be presented below, we take as radial function $u_{l,k_e}$  a regular  Coulomb function corresponding to a charge $z=1$. In an atomic or molecular ionization process this choice would describe a continuum electron ejected in a pure Coulomb potential, the charge being that felt asymptotically.
In this particular case, the integral~\eqref{eq:RadInteg_Idef} can be performed analytically~\cite{gradshteyn_2007}: the result, given in terms of a Gauss hypergeometric function  $_2F_1$, serves as a benchmark to validate other calculations.
In realistic atomic or molecular applications, one does not have a pure Coulomb potential, the ejected electron being then better described by a distorted wave \cite{ammar_2023} and for the corresponding numerical radial function $u_{l,k_e}$ such benchmark is not available. On the other hand, a \ac{cGTOs} representation can be used in the proposed integration scheme (see subsection~\ref{AnalyticalScheme}).

In physical applications $r$ stands for the radial distance, usually expressed in atomic units (a.u.).
However, since we are presenting a mathematical investigation, $r$ may be considered as a mathematical variable with no units.

\subsection{Difficulties related to the oscillating integrand}

\begin{figure}
	\begin{minipage}{\textwidth}
		\hspace{-3.90 cm}
		\begin{minipage}{0.50\textwidth}
			\includegraphics[scale=1.10]{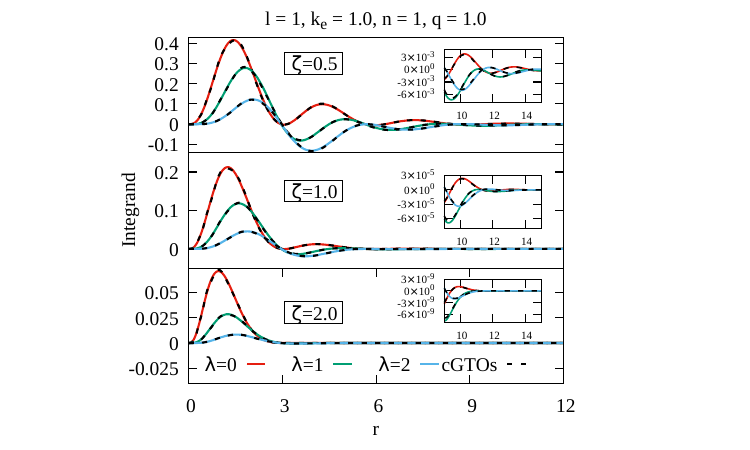}
		\end{minipage}
		\hfil
		\hspace{0.10 cm}
		\begin{minipage}{0.50\textwidth}
			\includegraphics[scale=1.10]{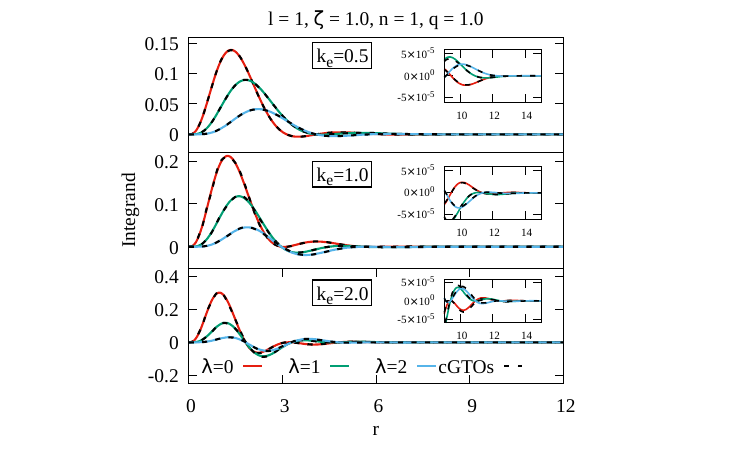}
		\end{minipage}
	\end{minipage}
	\vfill
	\begin{minipage}{\textwidth}
		\hspace{-3.90 cm}
		\begin{minipage}{0.50\textwidth}
			\includegraphics[scale=1.10]{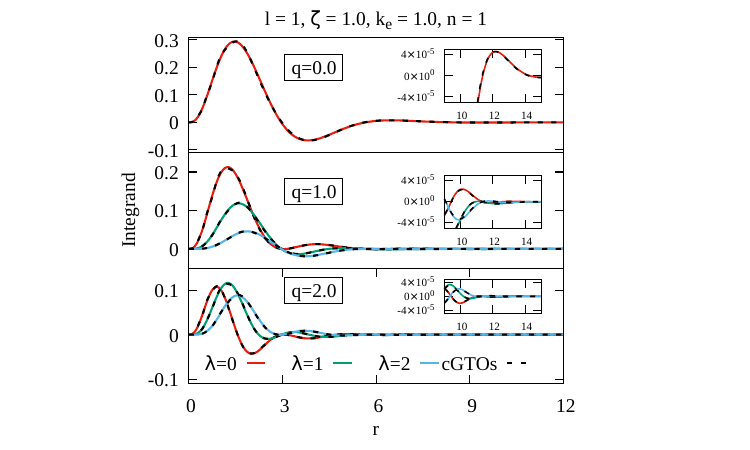}
		\end{minipage}
		\hfil
		\hspace{0.10 cm}
		\begin{minipage}{0.50\textwidth}
			\includegraphics[scale=1.10]{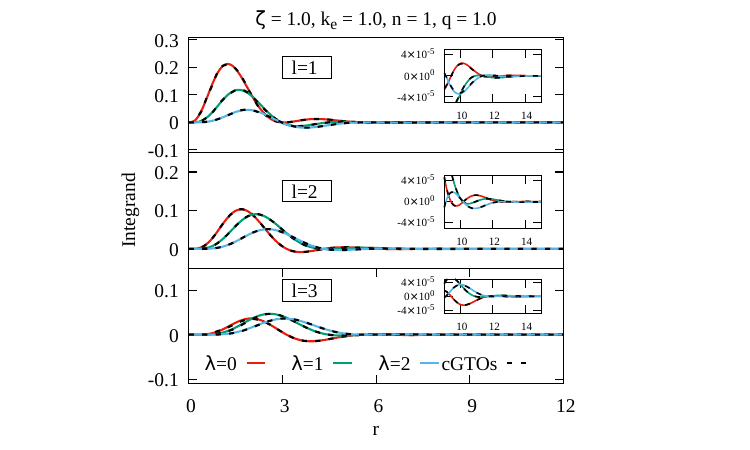}
		\end{minipage}
	\end{minipage}
	\caption{\label{fig:integrand_4sets}
		The integrand of~\eqref{eq:RadInteg_Idef} is plotted as a function of $r$
		for different sets of parameters $\{l,  \zeta, k_e, q\}$, $n$ being fixed to one.
		In each panel we fix three parameters and plot in three sub$-$panels the integrand for three different
		choices of the fourth parameter:
		the varying parameter is $\zeta$ in the top left panel, $k_e$ in the top right panel,
		$q$ in the bottom left panel and $l$ in the bottom right panel.
		In the 12 sub$-$panels, the integrand is
		plotted for $\lambda=0,1,2$. For each of the 36 situations,  the \ac{cGTOs} representation (sum of integrands in~\eqref{eq:RadInteg_IdefGTO})
		with $N=30$ is drawn in black dotted line.
	}
\end{figure}

Generally speaking, the evaluation of integrals~\eqref{eq:RadInteg_Idef} and~\eqref{eq:RadInteg_Jdef} can be challenging due to the oscillations in the integrand.
Their frequency, amplitude and range depend on the set of parameters $l,\zeta,k_e,n,q$ and $\lambda$.
To illustrate how these parameters act, we plot in Fig.~\ref{fig:integrand_4sets} the integrand of~\eqref{eq:RadInteg_Idef}
 for
different sets, $n$ being kept fixed to 1. Since the analysis of~\eqref{eq:RadInteg_Jdef} is quite similar, it will not be given here.
The values taken for the parameters are dictated by typical applications to low$-$energy ionisation processes~\cite{ammar_2021_I,ammar_2023}.
In the four panels of the figure, we fix all the parameters except one which is varied and presented in the three sub-panels. In each of the 12 sub-panels we plot the integrand for $\lambda=0,1,2$ and its  representation with $N=30$ \ac{cGTOs} (that is to say the sum of the integrands of equation~\eqref{eq:RadInteg_IdefGTO}).

The left top panel shows how the exponent $\zeta$ acts on the extension of the integrand. A small $\zeta$ leads to a larger effective integration range  and consequently the effective number of oscillations to be accounted for is increased.
The right top panel illustrates that the frequency of the oscillations is increased when the wave number $k_e$ is increased.
Note that, in the \ac{cGTOs} representation (\ref{eq:cG_def}), for each $k_e$ a different set of linear coefficients is optimized.
The left bottom panel shows a similar trend (increase of frequency)  but related to increasing values of $q$; contrary to the previous panel, $k_e$ is fixed and the same GTO expansion is used for the three $q$ values.
In the right bottom panel we can see that $l$ changes mainly the positions of the nodes and the amplitudes of the oscillations.
For each $l$, a different set of exponents and linear coefficients of \ac{cGTOs} are optimized.

The value of $\lambda$ - the index of the Bessel functions - also regulates the amplitudes and position of the oscillations (except for $q=0$).

The insets in the 12 sub-panels show how the integrand may oscillate up to large distances, albeit with rather small amplitudes. This indicates that the numerical evaluation of the corresponding integrals must be performed with care, since positive and negative contributions appear up to, in principle, infinity.


\begin{figure}
	\begin{minipage}{\textwidth}
		\center
		\includegraphics[scale=1.00]{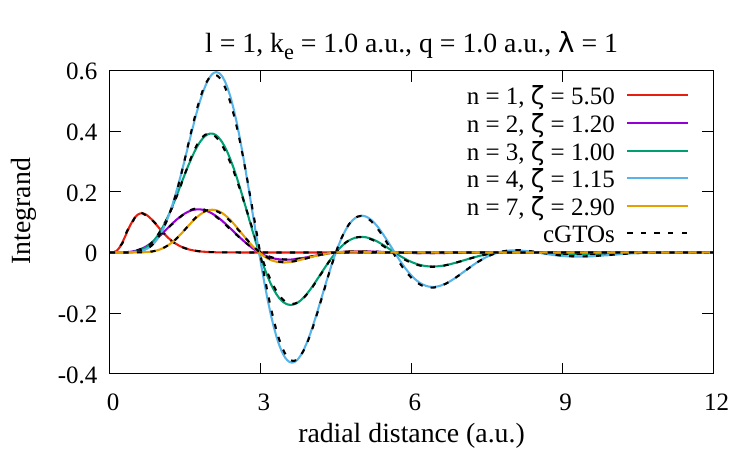}
	\end{minipage}
	\caption{\label{fig:integrand_MO}
		The integrand of~\eqref{eq:RadInteg_Idef} is plotted as a function of $r$
		for different sets of \ac{STOs} $\{n,\zeta\}$.
		These \ac{STOs} correspond to the most diffused functions among the molecular
		orbitals optimized in~\cite{moccia_1964_I,moccia_1964_II,moccia_1964_III} for each $n$.
		The red curve was multiplied by a factor of $100$.
		For each case the \ac{cGTOs} representation with $N=30$ is drawn in black dotted line.
	}
\end{figure}

For all the plotted curves, one observes that the \ac{cGTOs} provide a very good accurate fitting on the whole domain.
In physical applications, the range of the integrand of~\eqref{eq:RadInteg_Idef} and~\eqref{eq:RadInteg_Jdef} depends on the electronic extension of the molecule;
for small molecules, such as NH$_3$ or CH$_4$,  $r=30$ a.u. is generally more than sufficient. To show this, in Fig.~\ref{fig:integrand_MO}, the integrand of~\eqref{eq:RadInteg_Idef}
is plotted with realistic molecular orbitals taken from the literature, for fixed $l=1, k_e=1.0$ a.u., $q=1.0$ a.u. and $\lambda=1$.
For the present illustration we have selected, among the molecular orbitals optimized and tabulated by Moccia~\cite{moccia_1964_I,moccia_1964_II,moccia_1964_III},
five sets of $\{n,\zeta\}$ corresponding
to the most diffused \ac{STOs}
(\textit{i.e.}, the smallest exponent $\zeta$  for $n=1,2,3,4,7$). They are the most difficult to represent, and
the evaluation of the corresponding radial integrals is the most delicate.
Contrary to Fig.~\ref{fig:integrand_4sets},  here $n$ varies; the $r^n$ factor  provides at larger distances extra weight to the integrand.
In all five cases, the \ac{cGTOs} representation with $N=30$ reproduces very well the exact integrands.

\subsection{Reference value and error related to the range of integration}
\label{sec:HSPot}

All integrals $\mathcal{I}$~\eqref{eq:RadInteg_Idef} or $\mathcal{J}$~\eqref{eq:RadInteg_Jdef} can be evaluated with
a numerical quadrature, and the result will be labelled with the superscript \enquote{quad}.
To do so, we employ the \texttt{Fortran} library \texttt{QUADPACK}~\cite{piessens_2012}
where automatic routines are used to perform the integration with relative error tolerance of $10^{-12}$.
Since for the particular case of a regular Coulomb radial function the integral~\eqref{eq:RadInteg_IdefGTO} is known analytically, this numerical quadrature could be counterchecked.

As a first stage, we focus on the effective range of the integrals. Figures~\ref{fig:integrand_4sets} and \ref{fig:integrand_MO} showed that the integrand may oscillate up to large distances. Although the bound state finally makes the amplitude tend to zero, the value $\Rmax$ needed for the upper bound of the quadrature will determine the final accuracy of the integrals evaluation.
 To investigate the importance of the range of integration, we define the relative error
\begin{equation}
	\mathcal{E}_{\Rmax}
	= \left|
		\frac{\Iquad\left( \infty \right) - \Iquad\left( \Rmax \right)}{\Iquad\left( \infty \right)}
	\right|
	\text{,}
	\label{eq:error_Rmax_def}
\end{equation}
where $\Rmax = \infty$ corresponds to the
distance after which the integrand contribution is smaller than the tolerance error.
We also assume that $\Iquad\left( \infty \right)$ is sufficiently accurate as to be considered as the exact reference.

\begin{figure}
	\begin{minipage}{\textwidth}
		\hspace{-1.00 cm}
		\begin{minipage}{0.50\textwidth}
			\includegraphics[scale=0.65]{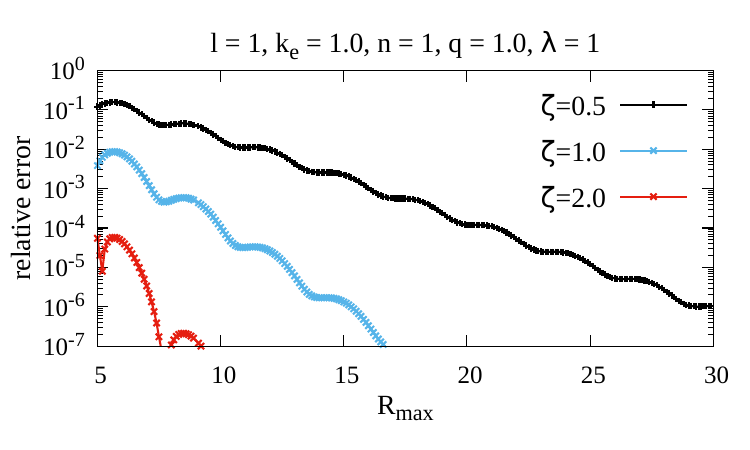}
		\end{minipage}
		\hfil
		\hspace{+1.00 cm}
		\begin{minipage}{0.50\textwidth}
			\includegraphics[scale=0.65]{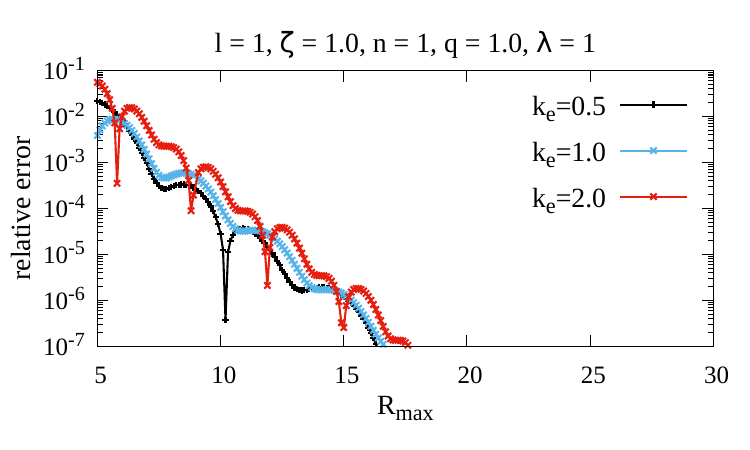}
		\end{minipage}
	\end{minipage}
	\vfill
	\vspace{0.10 cm}
	\begin{minipage}{\textwidth}
		\hspace{-1.00 cm}
		\begin{minipage}{0.50\textwidth}
			\includegraphics[scale=0.65]{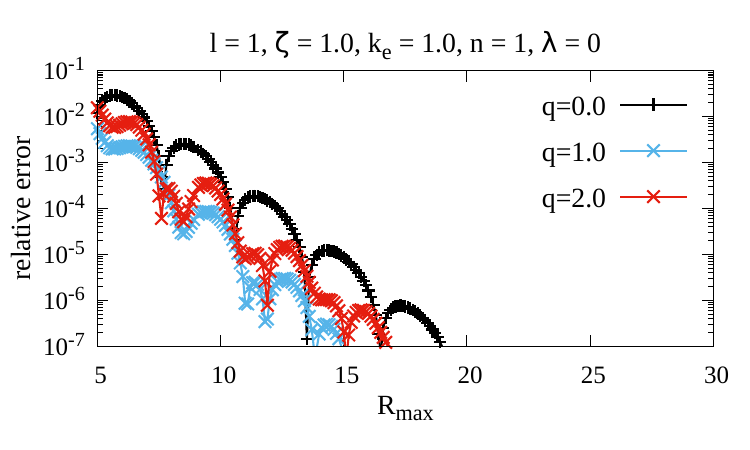}
		\end{minipage}
		\hfil
		\hspace{+1.00 cm}
		\begin{minipage}{0.50\textwidth}
			\includegraphics[scale=0.65]{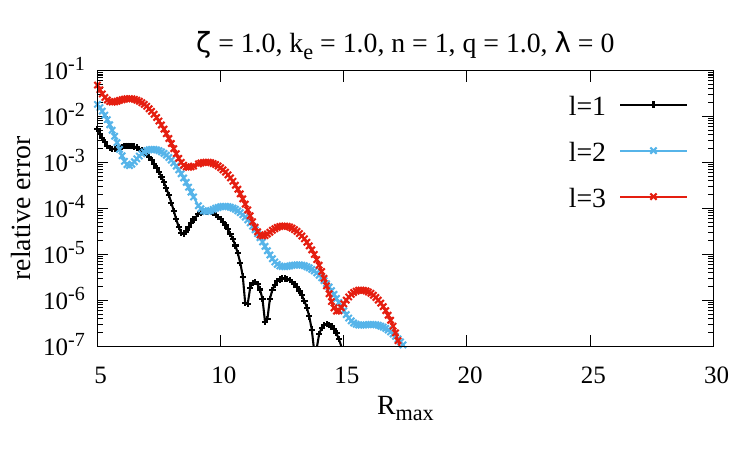}
		\end{minipage}
	\end{minipage}
	\caption{\label{fig:Err_Rmax}
		The error $\mathcal{E}_{\Rmax}$~\eqref{eq:error_Rmax_def} as a function of $\Rmax$ for different
		sets of parameters. Similarly to Figure~\ref{fig:integrand_4sets}, $n=1$ and in each panel we fix all the parameters except one.
	}
\end{figure}

The behaviour of the relative error $\mathcal{E}_{\Rmax}$ in terms of the upper bound $\Rmax$ is shown in Fig.~\ref{fig:Err_Rmax}.
To illustrate the impact of each parameter we present four panels in each of which only one parameter is changed. In all cases $n=1$ as in Figure~\ref{fig:integrand_4sets}.
One can see that, for all cases considered, $\Rmax=30$ is sufficient to achieve an acceptable accuracy (lower than $10^{-6}$).
This tells us that if we want to evaluate the integrals with the analytical approach  proposed in subsection~\ref{AnalyticalScheme},
 the \ac{cGTOs} representation must be reliable up to about $\Rmax=30$ (this is the radial domain we have considered here and in previous applications ~\cite{ammar_2020,ammar_2021_I,ammar_2021_II}).

\subsection{Comparison of integrals: numerical quadrature versus analytical \ac{cGTOs} approach}
\label{sec:Results}

We now wish to check the reliability of the \ac{cGTOs} representation to analytically evaluate the integrals
through expression (\ref{eq:I_analytical}) and (\ref{JKummer}).
%
For a quantitative comparison, we define the relative errors
\begin{align}
	\label{eq:error_cGTOs_STOs_def}
	\Err &= \left| \frac{\Iquad\left( \infty \right) - \IcGTOs}{\Iquad\left( \infty \right)} \right|
	\text{,} \\
	\label{eq:error_cGTOs_GTOs_def}
	\Frr &= \left| \frac{\Jquad\left( \infty \right) - \JcGTOs}{\Jquad\left( \infty \right)} \right|
	\text{,}
\end{align}
with respect to the reference values obtained by precise numerical quadrature.

In Tab.~\ref{tab:err_cGTOs_STOs}, we tabulate the values of the reference integrals~$\Iquad\left( \infty \right)$~\eqref{eq:RadInteg_Idef},
the \ac{cGTOs} integrals $\IcGTOs$
~\eqref{eq:I_analytical} with $N=30$,
and the relative error~\eqref{eq:error_cGTOs_STOs_def} for the 36 different sets of parameters considered in Fig.~\ref{fig:integrand_4sets}.
Although it is difficult to establish a strict dependence of these errors on each parameter,
we can make the following general observations:
$(i)$  by decreasing $\zeta$, the extension of the integrals becomes larger and so does the error;
 $(ii)$  the larger the wave number $k_e$, the faster the oscillations in $u_{l,k_e}$, and therefore the larger the error;
 $(iii)$ the variation of $q$ and $\lambda$ does not lead to any trend. While the change in
 $k_e$ has a direct impact on the quality of the \ac{cGTOs}, the change of $q$ and $\lambda$ does not affect them but influences
the oscillating nature of the integrand.

The relative errors vary between $0.004 \%$ and $5.6\%$.
We notice that amongst the 36 cases the largest errors appear for integrals whose absolute value is rather small.
Moreover, while individual integrals may not be perfectly evaluated, one has to take into account that
in applications they enter partial wave summations for which
rather rapid convergence is usually expected with respect to the $l$ and $\lambda$ values.

We performed a similar study by comparing the integrals  $\Jquad\left( \infty \right)$~\eqref{eq:RadInteg_Jdef} obtained by quadrature and the \ac{cGTOs} integrals 
~\eqref{JKummer} with $N=30$.
Tab.~\ref{tab:err_cGTOs_GTOs}
reports the relative errors ~\eqref{eq:error_cGTOs_GTOs_def} for four $\gamma$ values and five $n$ values, while keeping the other parameters fixed to
$l=1,k_e=1.0,q=1.0, \lambda=1$. The relative errors range from $0.03 \%$ to $5.2\%$.
Quite logically, larger relative errors are observed for smaller $\gamma$ values (more diffused bound states) and for larger $n$ values (enhanced weight on larger distances though the term $r^n$).


Finally we consider again the five sets of parameters $\{n,\zeta\}$ associated with the realistic molecular orbitals considered in Fig. \ref{fig:integrand_MO}.
In Tab.~\ref{tab:err_moccia} we tabulate the integrals~\eqref{eq:RadInteg_Idef} and the corresponding relative error~\eqref{eq:error_cGTOs_STOs_def}
generated by a \ac{cGTOs} representation
with $N=30$.
We observe that the error remains smaller than $3\%$, the worse case scenario being for $\lambda=2$ and $\{n=7,\zeta=2.90\}$.
 Here we should point out that in the tables provided by Moccia that molecular orbital, for example, is accompanied by a relatively rather small coefficient in the expansion (\ref{eq:MO_def}). Moreover, one has to recall that in physical applications the integrals with small value will have little weight in the overall calculation of matrix elements.


\subsection{Convergence with respect to the number of \ac{cGTOs}}
\label{subsec:ng}

\begin{figure}
	\center
	\includegraphics[scale=0.95]{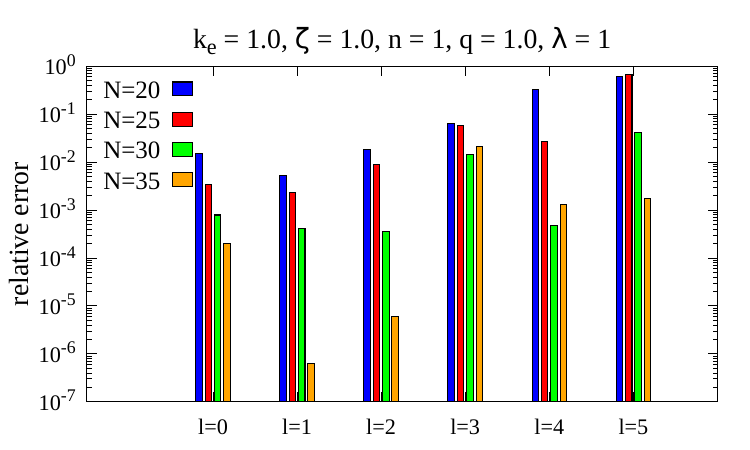}
	\caption{\label{fig:ng}
		The error $\Err$~\eqref{eq:error_cGTOs_STOs_def} for
		$N=20,25,30,35$ \ac{cGTOs}.
	}
\end{figure}

To close the analysis, we investigate the number of \ac{cGTOs} required for a good
representation of the continuum radial functions $u_{l,k_e}$ and the consequence on the integrals evaluation.
Fig.~\ref{fig:ng} shows the relative error $\Err$~\eqref{eq:error_cGTOs_STOs_def}  with $N=20,25,30,35$ \ac{cGTOs} for $l=0,\dots,5$ and for fixed parameters
$k_e=1,\zeta=1.0,n=1,q=1.0,\lambda=1$.
We recall here that, due to the non-linear \ac{cGTOs} optimization scheme \cite{ammar_2020}, the error corresponding to an energy $k_e$ does not decrease in a monotonic way with $N$. For the chosen parameters,
$N=30$ \ac{cGTOs} turns out to be a reasonable choice to get an acceptable accuracy
for angular momenta $l=0,\dots,5$. On the other hand $N=20$ and $N=25$ \ac{cGTOs} provide a good representation
only for small values of $l$. Going to $N=35$ \ac{cGTOs} improves slightly the quality.


\begin{landscape}

\begin{table}[]
{ \footnotesize
\center
\begin{tabular}{ccccccccccc}
	\toprule[1pt]
	\midrule[0.3pt]
	\multicolumn{1}{c|}{}               &
	\multicolumn{3}{|c|}{$\lambda = 0$} &
	\multicolumn{3}{|c|}{$\lambda = 1$} &
	\multicolumn{3}{|c}{ $\lambda = 2$} & \\
	\multicolumn{1}{c|}{}                                                                                              &
	\multicolumn{1}{|c|}{$\Iquad\left( \infty \right)$} & \multicolumn{1}{c|}{$\IcGTOs$} & \multicolumn{1}{c|}{$\Err$} &
	\multicolumn{1}{|c|}{$\Iquad\left( \infty \right)$} & \multicolumn{1}{c|}{$\IcGTOs$} & \multicolumn{1}{c|}{$\Err$} &
	\multicolumn{1}{|c|}{$\Iquad\left( \infty \right)$} & \multicolumn{1}{c|}{$\IcGTOs$} & \multicolumn{1}{c}{ $\Err$} \\
	\midrule[0.3pt]
 	\multicolumn{1}{c|}{\begin{tabular}{c} $\zeta = 0.5$\\ [\ExtraSep] $\zeta = 1.0$\\ [\ExtraSep] $\zeta = 2.0$ \end{tabular}} &
	\multicolumn{1}{|c}{ \begin{tabular}{c} +0.750188 \\[\ExtraSep] +0.305390 \\[\ExtraSep] +0.084562 \end{tabular}} &
	\multicolumn{1}{|c}{ \begin{tabular}{c} +0.750419 \\[\ExtraSep] +0.305281 \\[\ExtraSep] +0.084557 \end{tabular}} &
	\multicolumn{1}{|c|}{\begin{tabular}{c}  0.000308 \\[\ExtraSep]  0.000357 \\[\ExtraSep]  0.000064 \end{tabular}} &
	\multicolumn{1}{|c}{ \begin{tabular}{c} +0.292206 \\[\ExtraSep] +0.156200 \\[\ExtraSep] +0.036946 \end{tabular}} &
	\multicolumn{1}{|c}{ \begin{tabular}{c} +0.292448 \\[\ExtraSep] +0.156263 \\[\ExtraSep] +0.036943 \end{tabular}} &
	\multicolumn{1}{|c|}{\begin{tabular}{c}  0.000826 \\[\ExtraSep]  0.000403 \\[\ExtraSep]  0.000081 \end{tabular}} &
	\multicolumn{1}{|c}{ \begin{tabular}{c} -0.073052 \\[\ExtraSep] +0.035545 \\[\ExtraSep] +0.010671 \end{tabular}} &
	\multicolumn{1}{|c}{ \begin{tabular}{c} -0.073073 \\[\ExtraSep] +0.035687 \\[\ExtraSep] +0.010673 \end{tabular}} &
	\multicolumn{1}{|c}{ \begin{tabular}{c}  0.000297 \\[\ExtraSep]  0.004004 \\[\ExtraSep]  0.000200 \end{tabular}} & \\
	\midrule[0.3pt]
 	\multicolumn{1}{c|}{\begin{tabular}{c} $k_e = 0.5$\\ [\ExtraSep] $k_e = 1.0$\\ [\ExtraSep] $k_e = 2.0$ \end{tabular}} &
	\multicolumn{1}{|c}{ \begin{tabular}{c}  +0.212195 \\[\ExtraSep] +0.305390 \\[\ExtraSep] +0.229494  \end{tabular}} &
	\multicolumn{1}{|c}{ \begin{tabular}{c}  +0.212182 \\[\ExtraSep] +0.305281 \\[\ExtraSep] +0.229311  \end{tabular}} &
	\multicolumn{1}{|c|}{\begin{tabular}{c}   0.000062 \\[\ExtraSep]  0.000357 \\[\ExtraSep]  0.000795  \end{tabular}} &
	\multicolumn{1}{|c}{ \begin{tabular}{c}  +0.164112 \\[\ExtraSep] +0.156200 \\[\ExtraSep] +0.040133  \end{tabular}} &
	\multicolumn{1}{|c}{ \begin{tabular}{c}  +0.164131 \\[\ExtraSep] +0.156263 \\[\ExtraSep] +0.040334  \end{tabular}} &
	\multicolumn{1}{|c|}{\begin{tabular}{c}   0.000115 \\[\ExtraSep]  0.000403 \\[\ExtraSep]  0.005027  \end{tabular}} &
	\multicolumn{1}{|c}{ \begin{tabular}{c}  +0.075326 \\[\ExtraSep] +0.035545 \\[\ExtraSep] -0.005651  \end{tabular}} &
	\multicolumn{1}{|c}{ \begin{tabular}{c}  +0.075352 \\[\ExtraSep] +0.035687 \\[\ExtraSep] -0.005339  \end{tabular}} &
	\multicolumn{1}{|c}{ \begin{tabular}{c}   0.000344 \\[\ExtraSep]  0.004004 \\[\ExtraSep]  0.055282  \end{tabular}} & \\
	\midrule[0.3pt]
 	\multicolumn{1}{c|}{\begin{tabular}{c} $q = 0.0$\\ [\ExtraSep] $q = 1.0$\\ [\ExtraSep] $q = 2.0$ \end{tabular}} &
	\multicolumn{1}{|c}{ \begin{tabular}{c} +0.368802 \\[\ExtraSep] +0.305390  \\[\ExtraSep]  +0.049354  \end{tabular}} &
	\multicolumn{1}{|c}{ \begin{tabular}{c} +0.369971 \\[\ExtraSep] +0.305281  \\[\ExtraSep]  +0.049520  \end{tabular}} &
	\multicolumn{1}{|c|}{\begin{tabular}{c}  0.003172 \\[\ExtraSep]  0.000357  \\[\ExtraSep]   0.003354  \end{tabular}} &
	\multicolumn{1}{|c}{ \begin{tabular}{c} $-$       \\[\ExtraSep] +0.156200  \\[\ExtraSep]  +0.117574  \end{tabular}} &
	\multicolumn{1}{|c}{ \begin{tabular}{c} $-$       \\[\ExtraSep] +0.156263  \\[\ExtraSep]  +0.117350  \end{tabular}} &
	\multicolumn{1}{|c|}{\begin{tabular}{c} $-$       \\[\ExtraSep]  0.000403  \\[\ExtraSep]   0.001898  \end{tabular}} &
	\multicolumn{1}{|c}{ \begin{tabular}{c} $-$       \\[\ExtraSep] +0.035545  \\[\ExtraSep]  +0.112857 \end{tabular}} &
	\multicolumn{1}{|c}{ \begin{tabular}{c} $-$       \\[\ExtraSep] +0.035687  \\[\ExtraSep]  +0.112636 \end{tabular}} &
	\multicolumn{1}{|c}{ \begin{tabular}{c} $-$       \\[\ExtraSep]  0.004004  \\[\ExtraSep]   0.001964 \end{tabular}} & \\
	\midrule[0.3pt]
 	\multicolumn{1}{c|}{\begin{tabular}{c} $l = 1$\\ [\ExtraSep] $l = 2$\\ [\ExtraSep] $l = 3$ \end{tabular}} &
	\multicolumn{1}{|c}{ \begin{tabular}{c} +0.305390  \\[\ExtraSep] +0.151588  \\[\ExtraSep] +0.032533  \end{tabular}} &
	\multicolumn{1}{|c}{ \begin{tabular}{c} +0.305281  \\[\ExtraSep] +0.151733  \\[\ExtraSep] +0.030699  \end{tabular}} &
	\multicolumn{1}{|c|}{\begin{tabular}{c}  0.000357  \\[\ExtraSep]  0.000959  \\[\ExtraSep]  0.056352  \end{tabular}} &
	\multicolumn{1}{|c}{ \begin{tabular}{c} +0.156200 \\[\ExtraSep] +0.165232  \\[\ExtraSep]  +0.088876 \end{tabular}} &
	\multicolumn{1}{|c}{ \begin{tabular}{c} +0.156263 \\[\ExtraSep] +0.165292  \\[\ExtraSep]  +0.087595 \end{tabular}} &
	\multicolumn{1}{|c|}{\begin{tabular}{c}  0.000403 \\[\ExtraSep]  0.000361  \\[\ExtraSep]   0.014413 \end{tabular}} &
	\multicolumn{1}{|c}{\begin{tabular}{c} +0.035545  \\[\ExtraSep] +0.092021  \\[\ExtraSep] +0.078056  \end{tabular}} &
	\multicolumn{1}{|c}{\begin{tabular}{c} +0.035687  \\[\ExtraSep] +0.092017  \\[\ExtraSep] +0.077610  \end{tabular}} &
	\multicolumn{1}{|c}{\begin{tabular}{c}  0.004004  \\[\ExtraSep]  0.000042  \\[\ExtraSep]  0.005711  \end{tabular}} &  \\
	\midrule[0.3pt]
	\toprule[1pt]
\end{tabular}
}
	\caption{\label{tab:err_cGTOs_STOs}
		Reference integrals $\Iquad\left( \infty \right)$~\eqref{eq:RadInteg_Idef},
		 \ac{cGTOs} integrals $\IcGTOs$
~\eqref{eq:I_analytical} with $N=30$,
		and the relative error~\eqref{eq:error_cGTOs_STOs_def} for the 36 sets of parameters
		considered in Fig.~\ref{fig:integrand_4sets}.
		We fix
		$\{   l = 1,                k_e = 1.0, n = 1,     q = 1.0 \}$,
		$\{   l = 1,   \zeta = 1.0,            n = 1,     q = 1.0 \}$,
		$\{   l = 1,   \zeta = 1.0, k_e = 1.0, n = 1 \}$,
		$\{            \zeta = 1.0, k_e = 1.0, n = 1,     q = 1.0 \}$,
		in the first, second, third and last block of three lines , respectively.
	}
\end{table}
\end{landscape}

\begin{table}[]
{ \footnotesize
\center
\begin{tabular}{ccccccc}
	\toprule[1pt]
	\midrule[0.3pt]
	\multicolumn{1}{c}{}                                &
	\multicolumn{1}{c|}{}                               &
	\multicolumn{1}{|c|}{$\Jquad\left( \infty \right)$} &
	\multicolumn{1}{|c|}{$\JcGTOs$}                     &
	\multicolumn{1}{|c}{ $\Frr$}                        & \\
	\midrule[0.3pt]
	\multicolumn{1}{c}{$\gamma=0.01$} &
 	\multicolumn{1}{c|}{\begin{tabular}{c} $n = 1$\\ [\ExtraSep] $n=2$ \\[\ExtraSep] $n=3$\\ [\ExtraSep] $n=4$\\ [\ExtraSep] $n=5$ \end{tabular}} &
	\multicolumn{1}{|c|}{\begin{tabular}{c}-0.782320\\[\ExtraSep]-13.73538\\[\ExtraSep]-164.6196\\[\ExtraSep]-2100.598\\[\ExtraSep]-29673.28\\[\ExtraSep]\end{tabular}} &
	\multicolumn{1}{|c|}{\begin{tabular}{c}-0.765971\\[\ExtraSep]-13.56619\\[\ExtraSep]-163.0429\\[\ExtraSep]-2086.769\\[\ExtraSep]-29524.63\\[\ExtraSep]\end{tabular}} &
	\multicolumn{1}{|c}{ \begin{tabular}{c} 0.020899\\[\ExtraSep]  0.012318\\[\ExtraSep]   0.009578\\[\ExtraSep]    0.006583\\[\ExtraSep]     0.005010\\[\ExtraSep]\end{tabular}} & \\
	\midrule[0.3pt]
	\multicolumn{1}{c}{$\gamma=0.05$} &
 	\multicolumn{1}{c|}{\begin{tabular}{c} $n = 1$\\ [\ExtraSep] $n=2$ \\[\ExtraSep] $n=3$\\ [\ExtraSep] $n=4$\\ [\ExtraSep] $n=5$ \end{tabular}} &
	\multicolumn{1}{|c|}{\begin{tabular}{c} +0.529978\\[\ExtraSep] +0.182134\\[\ExtraSep] -4.084013\\[\ExtraSep] -37.21119\\[\ExtraSep]-287.6302\\[\ExtraSep]\end{tabular}} &
	\multicolumn{1}{|c|}{\begin{tabular}{c} +0.529091\\[\ExtraSep] +0.172596\\[\ExtraSep] -4.133912\\[\ExtraSep] -37.25801\\[\ExtraSep]-284.9418\\[\ExtraSep]\end{tabular}} &
	\multicolumn{1}{|c}{ \begin{tabular}{c}  0.001674\\[\ExtraSep]  0.052367\\[\ExtraSep]  0.012218\\[\ExtraSep]   0.001258\\[\ExtraSep]   0.009347\\[\ExtraSep]\end{tabular}} & \\
	\midrule[0.3pt]
	\multicolumn{1}{c}{$\gamma=0.1$} &
 	\multicolumn{1}{c|}{\begin{tabular}{c} $n = 1$\\ [\ExtraSep] $n=2$ \\[\ExtraSep] $n=3$\\ [\ExtraSep] $n=4$\\ [\ExtraSep] $n=5$ \end{tabular}} &
	\multicolumn{1}{|c|}{\begin{tabular}{c} +0.567314\\[\ExtraSep] +0.803623\\[\ExtraSep] +0.714211\\[\ExtraSep] -2.183815\\[\ExtraSep] -22.73327\\[\ExtraSep] \end{tabular}} &
	\multicolumn{1}{|c|}{\begin{tabular}{c} +0.568264\\[\ExtraSep] +0.805258\\[\ExtraSep] +0.706394\\[\ExtraSep] -2.294839\\[\ExtraSep] -23.58571\\[\ExtraSep] \end{tabular}} &
	\multicolumn{1}{|c}{ \begin{tabular}{c}  0.001675\\[\ExtraSep]  0.002036\\[\ExtraSep]  0.010945\\[\ExtraSep]  0.050839\\[\ExtraSep]   0.037497\\[\ExtraSep] \end{tabular}} & \\
	\midrule[0.3pt]
	\multicolumn{1}{c}{$\gamma=0.5$} &
 	\multicolumn{1}{c|}{\begin{tabular}{c} $n = 1$\\ [\ExtraSep] $n=2$ \\[\ExtraSep] $n=3$\\ [\ExtraSep] $n=4$\\ [\ExtraSep] $n=5$ \end{tabular}} &
	\multicolumn{1}{|c|}{\begin{tabular}{c} +0.208727\\[\ExtraSep] +0.310880\\[\ExtraSep] +0.505495\\[\ExtraSep] +0.877332\\[\ExtraSep] +1.593688\\[\ExtraSep] \end{tabular}} &
	\multicolumn{1}{|c|}{\begin{tabular}{c} +0.208658\\[\ExtraSep] +0.310665\\[\ExtraSep] +0.504953\\[\ExtraSep] +0.876267\\[\ExtraSep] +1.592594\\[\ExtraSep] \end{tabular}} &
	\multicolumn{1}{|c}{ \begin{tabular}{c}  0.000330\\[\ExtraSep]  0.000692\\[\ExtraSep]  0.001073\\[\ExtraSep]  0.001213\\[\ExtraSep]  0.000686\\[\ExtraSep] \end{tabular}} & \\
	\midrule[0.3pt]
	\toprule[1pt]
\end{tabular}
}
	\caption{\label{tab:err_cGTOs_GTOs}
		Reference integrals $\Jquad\left( \infty \right)$~\eqref{eq:RadInteg_Jdef},
		\ac{cGTOs} integrals
~\eqref{JKummer}
with $N=30$,
		and the relative error~\eqref{eq:error_cGTOs_GTOs_def} for different sets of parameters $n$ and $\gamma$.
		The results are obtained for $l=1,k_e=1.0,q=1.0, \lambda=1$.
	}
\end{table}

\begin{landscape}

\begin{table}[]
{ \footnotesize
\center
\begin{tabular}{ccccccccccc}
	\toprule[1pt]
	\midrule[0.3pt]
	\multicolumn{1}{c|}{}               &
	\multicolumn{3}{|c|}{$\lambda = 0$} &
	\multicolumn{3}{|c|}{$\lambda = 1$} &
	\multicolumn{3}{|c}{ $\lambda = 2$} & \\
	\multicolumn{1}{c|}{}                                                                                              &
	\multicolumn{1}{|c|}{$\Iquad\left( \infty \right)$} & \multicolumn{1}{c|}{$\IcGTOs$} & \multicolumn{1}{c|}{$\Err$} &
	\multicolumn{1}{|c|}{$\Iquad\left( \infty \right)$} & \multicolumn{1}{c|}{$\IcGTOs$} & \multicolumn{1}{c|}{$\Err$} &
	\multicolumn{1}{|c|}{$\Iquad\left( \infty \right)$} & \multicolumn{1}{c|}{$\IcGTOs$} & \multicolumn{1}{c}{ $\Err$} \\
	\midrule[0.3pt]
 	\multicolumn{1}{c|}{\begin{tabular}{c} $n = 1$ \\[\ExtraSep] $\zeta = 5.50$ \end{tabular}} &
	\multicolumn{1}{|c}{ +0.0046639 } &
	\multicolumn{1}{|c}{ +0.0046643 } &
	\multicolumn{1}{|c|}{ 0.000079 } &
	\multicolumn{1}{|c}{ +0.001005011 } &
	\multicolumn{1}{|c}{ +0.001005014 } &
	\multicolumn{1}{|c|}{ 0.000003 } &
	\multicolumn{1}{|c}{ +0.0001594588 } &
	\multicolumn{1}{|c}{ +0.0001594594 } &
	\multicolumn{1}{|c}{  0.000004 } & \\
	\midrule[0.3pt]
 	\multicolumn{1}{c|}{\begin{tabular}{c} $n = 2$ \\[\ExtraSep] $\zeta = 1.20$ \end{tabular}} &
	\multicolumn{1}{|c} { +0.318374 } &
	\multicolumn{1}{|c} { +0.318202 } &
	\multicolumn{1}{|c|}{ 0.000543  } &
	\multicolumn{1}{|c} { +0.172295 } &
	\multicolumn{1}{|c} { +0.172431 } &
	\multicolumn{1}{|c|}{ 0.000789  } &
	\multicolumn{1}{|c} { +0.030050 } &
	\multicolumn{1}{|c} { +0.030309 } &
	\multicolumn{1}{|c} {  0.008624 } & \\
	\midrule[0.3pt]
 	\multicolumn{1}{c|}{\begin{tabular}{c} $n = 3$ \\[\ExtraSep] $\zeta = 1.00$ \end{tabular}} &
	\multicolumn{1}{|c} { +0.925984  } &
	\multicolumn{1}{|c} { +0.926280  } &
	\multicolumn{1}{|c|}{  0.000319  } &
	\multicolumn{1}{|c} { +0.312400  } &
	\multicolumn{1}{|c} { +0.312567  } &
	\multicolumn{1}{|c|}{  0.000533  } &
	\multicolumn{1}{|c} { -0.235702  } &
	\multicolumn{1}{|c} { -0.235400  } &
	\multicolumn{1}{|c} {  0.001280  } & \\
	\midrule[0.3pt]
 	\multicolumn{1}{c|}{\begin{tabular}{c} $n = 4$ \\[\ExtraSep] $\zeta = 1.15$ \end{tabular}} &
	\multicolumn{1}{|c} { +1.512731  } &
	\multicolumn{1}{|c} { +1.514143  } &
	\multicolumn{1}{|c|}{  0.000933  } &
	\multicolumn{1}{|c} { +0.310337 } &
	\multicolumn{1}{|c} { +0.309939 } &
	\multicolumn{1}{|c|}{  0.001283 } &
	\multicolumn{1}{|c} { -0.686099  } &
	\multicolumn{1}{|c} { -0.686402  } &
	\multicolumn{1}{|c} {  0.000442  } & \\
	\midrule[0.3pt]
 	\multicolumn{1}{c|}{\begin{tabular}{c} $n = 7$ \\[\ExtraSep] $\zeta = 2.90$ \end{tabular}} &
	\multicolumn{1}{|c} { +0.206853  } &
	\multicolumn{1}{|c} { +0.206441  } &
	\multicolumn{1}{|c|}{  0.001990  } &
	\multicolumn{1}{|c} { +0.132105  } &
	\multicolumn{1}{|c} { +0.132482  } &
	\multicolumn{1}{|c|}{  0.002849  } &
	\multicolumn{1}{|c} { +0.021776  } &
	\multicolumn{1}{|c} { +0.022351  } &
	\multicolumn{1}{|c} {  0.026389  } & \\
	\midrule[0.3pt]
	\toprule[1pt]
\end{tabular}
}
	\caption{\label{tab:err_moccia}
		Reference integrals $\Iquad\left( \infty \right)$~\eqref{eq:RadInteg_Idef},
		\ac{cGTOs} integrals $\IcGTOs$
~\eqref{eq:I_analytical} with $N=30$,
		and the relative error~\eqref{eq:error_cGTOs_STOs_def} for different set of realistic \ac{STOs}
		$\{n,\zeta\}$~\cite{moccia_1964_I,moccia_1964_II,moccia_1964_III}
		where $l=1$, $k_e=1.0$, $q=1.0$ and for $\lambda=0,1,2$.
	}
\end{table}
\end{landscape}


\section{Summary}
\label{sec:Summary}
We have presented an investigation of radial integrals involving product of oscillating functions,  powers and decreasing exponentials.
Two classes are considered corresponding, respectively, to a STO or GTO description of bound states. The oscillating functions are related to an electron in a continuum state that arises, for example, after an atomic or molecular ionization process.
The present study illustrated some of the difficulties one may encounter in the evaluation of such integrals. Moreover, using a representation of the continuum radial function in terms of complex GTO, these integrals may be performed analytically and expressed as a sum of special functions.
The efficiency and limitation of such an approach are investigated for a range of the integrands' parameters.
 In all the tested cases, and even for realistic orbital parameters, the analytical integrals based on approximate cGTO expansions lead to accuracies that are always sufficient for physical applications in the low-energy domain \cite{ammar_2021_I,ammar_2023}.

The present work aimed to grasp the potential of the proposed approach by exploring its reliability for the monocentric case. We consider this step necessary
before tackling more difficult integrals, such as two-electron integrals or one electron integrals involving multicentric functions.
The ultimate goal would be to reach an all-Gaussian approach to evaluate the necessary integrals involved in scattering processes, similarly to the well known and widely used Gaussian tools and packages used in quantum chemistry.








\bibliographystyle{elsarticle-num}

\providecommand{\noopsort}[1]{}\providecommand{\singleletter}[1]{#1}%


%

\end{document}